\documentstyle[aps,pre,multicol,amsmath,epsf,graphics,epsfig]{revtex}

\tighten

\begin{document}

\title{Ferromagnetic ordering in graphs with arbitrary degree distribution}

\author{$^1$Michele Leone, $^1$Alexei V\'azquez, $^2$Alessandro Vespignani 
and $^2$Riccardo Zecchina}

\address{$^1$International School for Advanced Studies and INFM, via
Beirut 4, 34014 Trieste, Italy}

\address{$^2$The Abdus Salam International Centre for Theoretical
Physics, P.O. Box 586, 34014 Trieste, Italy}                                    

\date{\today}

\maketitle

\begin{abstract}

We present a detailed study of the phase diagram of the Ising model in random graphs with arbitrary
degree distribution. By using the replica method we compute exactly the value of the critical
temperature and the associated critical exponents as a function of the minimum and maximum 
degree, and the degree
distribution characterizing the graph. As expected, there is a ferromagnetic transition provided
$\left<k\right>\leq2\left<k^2\right><\infty$. However, if the fourth moment of the degree
distribution is not finite then non-trivial scaling exponents are obtained. These results are
analyzed for the particular case of power-law distributed random graphs.


\end{abstract}

\section{Introduction}

The increasing evidence that many physical, biological and social networks exhibit a high degree of
wiring entanglement has led to the investigation of graph models with complex topological
properties\cite{strogatz01}. In particular, the possibility that some special nodes of the cluster
(hubs) posses a larger probability to develop connections pointing to other nodes has been recently
identified in scale-free networks \cite{barabasi01,dorogovtsev01}. These networks exhibit a power law
degree distribution $p_k\sim k^{-\gamma}$, where the exponent $\gamma$ is usually larger than $2$.
This kind of degree distribution implies that each node has a statistically significant probability of
having a large number of connections compared to the average degree of the network.  Examples of
such properties can be found in communication and social webs, along with many biological networks,
and have led to the developing of several dynamical models aimed to the description and
characterization of scale-free networks\cite{barabasi01,dorogovtsev01,barabasi99}.

Power law degree distributions are the signature of degree fluctuations that may alter the phase
diagram of physical processes as in the case of random percolation \cite{havlin01,callaway01} and
spreading processes \cite{vespignani01} that do not exhibit a phase transition if the degree exponent
is $\gamma\leq 3$.  In this perspective, it is interesting to study the ordering dynamics of the
Ising model in scale-free networks. The Ising model is, indeed, the prototypical model for the study
of phase transitions and complex phenomena and it is often the starting point for the developing of
models aimed at the characterization of ordering phenomena. For this reason, the Ising model and its
variations are used to mimic a wide range of phenomena not pertaining to physics, such as the forming
and spreading of opinions in societies and companies or the evolution and competition of species.  
Since social and biological networks are often characterized by scale-free properties, the study of
the ferromagnetic phase transition in graphs with arbitrary degree distribution can find useful
application in the study of several complex interacting systems and it has been recently pursued in
Ref.~\cite{aleksiejuk01}. The numerical simulations reported in Ref~\cite{aleksiejuk01} show that in
the case of a degree distribution with $\gamma=3$ the Ising model has a critical temperature $T_c$,
characterizing the transition to an ordered phase, which scales logarithmically with the network
size. Therefore, there is no ferromagnetic transition in the thermodynamic limit.

In the present paper we present a detailed analytical study of the Ising model in graphs with
arbitrary degree distribution.  By relaxing the degree homogeneity in the usual mean field (MF)
approach to the Ising model, it is possible to show that the existence of a disordered phase is
related to the ratio of the first two moments of the degree distribution. Motivated by this finding,
we apply the replica calculation method, as developed for spin glasses and diluted ferromagnetic
models on random graphs ~\cite{MOZE,DILUTED,MEPA,RIWEZE,2p,lungo,FMRWZ},in order to find an exact
characterization of the transition to the ordered state and its associated critical behavior.  We
find that a disordered phase is allowed only if the second moment of the degree distribution is
finite. In the opposite case, the strong degree of the hubs present in the network prevails on the
thermal fluctuations, imposing a long-range magnetic order for any finite value of the temperature.
Corrections to this picture are found when the minimal allowed degree is $m=1$. The value of the
critical temperature and exponents is found for any degree exponent $\gamma>3$ and a transition to
the usual infinite dimensional MF behavior is recovered at $\gamma=5$. Moreover, in the range
$3<\gamma\leq5$ non trivial scaling exponents are obtained.
 
During the completion of the present work we become aware that Dorogovtsev, Goltsev and Mendes
\cite{doroising} have obtained with a different approach results which partially overlap 
with those reported in the present paper.

\section{Simple mean field approach}

Let us consider a network with arbitrary degree distribution $p_k$. Then consider the Ising model with
a ferromagnetic coupling constant on top of this network. The
Hamiltonian of this system is given by
\begin{equation}
{\cal H}=M-\sum_{i>j=1}^NJ_{ij}s_is_j + H_0 \sum_i \zeta_i s_i,
\label{eq:0}
\end{equation} 
where $M=\left<k\right>N/2$, $J_{ij}=1(0)$ if there is (there is not) a link connecting node $i$ and
$j$, $s_i=\pm1$ are the spin variables, and $N$ is the network size. $H_0 \zeta_i$ is a general external
random field with $\zeta_i $ following the a priori general probability distribution $\Lambda (\zeta_i)$.

Let us firstly put the external field $H_0 = 0$.
In the usual MF approximation one is able to give a first estimate of the critical temperature
and the magnetization that already takes into account the inhomogeneity of the graph. 
Neglecting the spin-spin correlations, $\left<s_i
s_j\right>\simeq\left<s_i\right>\left<s_j\right> =\left<s\right>^2$, it is then possible to use an
effective field ansatz in which each spin feels the average magnetization on neighboring spins
obtaining
\begin{equation}
{\cal H}_{MF}=M-\left<s\right>\sum_i k_i s_i,
\label{eq:001}
\end{equation}
where $k_i=\sum_jc_{ij}$ is the node degree. In the case of a spin network with homogeneous
degree $k_i=<k>$ and the average magnetization is found self-consistently obtaining
$\left<s\right>=\tanh\left(\beta <k>\left<s\right>\right)$ where $\beta$ is the inverse temperature in
units of $k_B^{-1}$.

In the case of complex heterogeneous networks, we can relax the homogeneity assumption on the node's
degree by defining the average magnetization $\left<s\right>_k$ for the class of spins with degree
$k$. Indeed, the node's magnetization is strongly affected by the local degree and the homogeneity
assumption results to be too drastic especially in singular degree distribution.  The self-consistent
equation for the average magnetization in each degree class reads simply as
$\left<s\right>_k=\tanh\left(\beta k\left<u\right>\right)$, where $\left<u\right>$ is now the
effective field magnetization seen by each node on the nearest neighbors. In the calculation of the
effective field we have to take into account the system's heterogeneity by noticing that each link
points to more likely to nodes with higher degree.  In particular the probability that a link points
to a node $j$ with degree $k_j$ is $k_j/\sum_lk_l$. Thus, the correct average magnetization seen on a
nearest neighbor node is given by
\begin{equation}
\left<u\right>=\frac{\sum_kkp_k\tanh\left(\beta k\left<u\right>\right)}
{<k>}.
\label{eq:4}
\end{equation}
Once obtained $\left<u\right>$ it is possible to compute the network average magnetization as the
average over all the degree classes $\left<s\right>=\sum_kp_k\tanh\left(\beta
k\left<u\right>\right)$. A non-zero magnetization solution is obtained when the tangent curves of the
lhs as a function of $\left<u\right>$ is greater than one and when the tangent is strictly equal to
one we obtain the critical point that defines
\begin{equation}
T_c=\beta_c^{-1}=\frac{\left<k^2\right>}{\left<k\right>}.
\label{eq:5}
\end{equation}
Hence, when $\left<k^2\right>/\left<k\right>$ is finite there is a finite critical temperature as an
evidence of the transition from the para-magnetic to a ferro-magnetic state. On the contrary, if
$\left<k^2\right>$ is not finite the system is always in the ferromagnetic state.
 
\section{The replica approach on general random graphs}
 
In the present section we will refine the mean field picture via a replica calculation, in the
framework of the method applied in the last years for general spin glasses and diluted ferromagnetic
models on random graphs. We will show briefly how this method allows to calculate values of and
conditions for the existence of a critical temperature of the model that we believe to be exact.
Moreover, these results contain the classical mean field theory prediction in the limits where the
latter is applicable.

For a random graph the interaction matrix elements in Eq. (\ref{eq:0}) follow the distribution
$P(J_{ij}) =  (1 - \frac{2}{N}) \delta (J_{ij}) + \frac{2}{N} \delta (J_{ij} -1)$
with constraints in order to impose the correct degree distribution that will have to be
introduced along the computation of the logarithm of the partition function. Following the approach of
Ref.~\cite{RIWEZE}, we compute the free energy of the model with the replica method, exploiting the
identity $\log <Z^n> = 1 + n <\log Z> + O(n^2)$. A part from a normalization factor:
\begin{equation} <Z^n> = e^{-n \beta M} \sum_{\vec{s_i}}
<e^{\beta \sum_{a=1}^{n} \sum_{i<j} J_{ij} s_{i}^a s_{j}^a} <e^{\beta
H_0\sum_{a=1}^n\sum_i
\zeta_i s_i^a}>_{\zeta_i}>
\end{equation} 
where $<...>_{\zeta_i}$ is the average over the quenched field and $<...>$ that 
over the quenched disorder:
$<A> = \int \prod_{i<j} d J_{ij} P(J_{ij}) \prod_{i=1}^{N} \delta \left(\sum_{<j>_i} J_{ij} - k_i\right) \, A $.
The delta functions are there to enforce the satisfaction of the constraints on the connectivities
distribution.  Writing the constraints in integral form
\begin{equation} 
\prod_i \delta
\left(\sum_{<j>_i} J_{ij} - k_i\right) = \int \prod_i \left(\frac{d {\psi}_i}{2 \pi}\right)
\exp \left(-i \sum_i {\psi}_i k_i\right) \exp \left(i \sum_{i<j} ({\psi}_i + {\psi}_j)
J_{ij}\right)
\end{equation}
we can write  
\begin{eqnarray} <Z^n> &\sim&  \exp
(-\beta n M) \sum_{\vec s_i} \int \prod_i \left(\frac{d {\psi}_i}{2 \pi}\right)
\exp \left(-i \sum_{i=1}^N \psi_i k_i\right) \nonumber \\
& &  \exp \left[-\frac{<k>}{2} N +
\frac{1}{N} \exp \left(\beta \sum_a s_{i}^as_{j}^a +  i (\psi_{i} + \psi_{j})\right)\right]
<e^{\beta H_0\sum_{a=1}^n\sum_i \zeta_i s_i^a}>_{\zeta_i} 
\end{eqnarray} 

We now introduce a functional order parameter, following the well tested procedure of replica theory
of diluted systems~\cite{DILUTED}:
\begin{equation} \rho (\vec{\sigma}) =
\frac{1}{N} \sum_i \delta (\vec{\sigma} - {\vec{s}}_i) e^{i \psi_i}
\label{eq:order_par} 
\end{equation} 
Tracing over the spins $\vec{s}_i$, integrating out the ${\psi}_i$ variables and introducing the
correct normalization factor for the constrained probabilities $P(J_{ij})$ one is left with the
following expression for the replica free energy:
\begin{eqnarray} 
-n \beta F &=& -<k> \sum_{\vec{\sigma}}
\rho (\vec{\sigma}) \hat{\rho}(\vec{\sigma}) + \frac{<k>}{2}\left[1-n \beta
+ \sum_{{\vec{\sigma}}_1,{\vec{\sigma}}_2} \rho ({\vec{\sigma}}_1)
\rho ({\vec{\sigma}}_2) \exp \left(\beta \sum_a {\sigma}_1^a {\sigma}_2^a\right)\right]
+ \nonumber \\
& & \sum_k p_k \log \left[\sum_{\vec{\sigma}} (\hat{\rho
(\vec{\sigma})})^{k}H(\vec{\sigma})\right] \label{eq:potenziale} 
\end{eqnarray} 
where $\hat{\rho}(\vec{\sigma})$ is the functional order parameter
conjugate to $\rho(\vec{\sigma})$ and $H(\vec{\sigma}) = <e^{H_0 \zeta
\sum_a \sigma^a}>_{\zeta} = \int d \zeta \Lambda (\zeta)  e^{H_0 \zeta
\sum_a \sigma^a}$. In the following we will concentrate on the fixed
external field case $\Lambda (\zeta) = \delta (\zeta - 1)$, being the
study of the model over a true random field beyond the scope of this
work. We'll treat the general case  in a forthcoming article \cite{lungo}.  
The main contribution to the free energy in the thermodynamic limit is
evaluated via the following functional saddle point equations:
\begin{eqnarray} 
\rho (\vec{\sigma}) &=&
\frac{1}{<k>} \sum_k k p_k \frac{ (\hat{\rho} (\vec{\sigma}))^{k-1} H(\vec{\sigma})}
{\sum_{\vec{\sigma}} (\hat{\rho} (\vec{\sigma}))^{k}H(\vec{\sigma})}
\label{eq:saddle1a} \\ \hat{\rho} (\vec{\sigma}) &=&
\sum_{{\vec{\sigma}}_1} \rho ({\vec{\sigma}}_1) \exp (\beta \sum_a
{\sigma}^a {\sigma}_1^a)  \label{eq:saddle1b} 
\end{eqnarray} 
It is easy to show that the order parameters can be taken as normalized in the $n \to 0$ limit.
Further normalization factors cancel out in the expression for the free energy.

\subsection{Solution of the saddle point equations}
\label{sec:sadle}

Being the system a diluted ferromagnet the replica symmetric ansatz is sufficient to find the 
correct solution of (\ref{eq:saddle1a}) and (\ref{eq:saddle1b}) at every temperature. In the general
case we can write:
\begin{eqnarray}
   \rho (\vec{\sigma}) &=& \int d h P(h)  \frac{e^{\beta h
   \sum_{a=1}^{n} \sigma_a}}{(2 \cosh (\beta h))^n} \label{eff1} \\ \hat{\rho}
   (\vec{\sigma}) &=& \int d u Q(u)  \frac{e^{\beta u \sum_{a=1}^{n}
   \sigma_a}}{(2 \cosh (\beta u))^n} \label{eff2}
\end{eqnarray}   
that leads to
\begin{eqnarray}
   P(h) &=& \frac{1}{<k>}  \sum_k k p_k  \int \prod_{t=1}^{k-1} d u_t
   Q(u_t) \delta \left(h - \sum_t u_t - H_0\right) \\  Q(u) &=& \int d h P(h) \delta \left[u
   - \frac{1}{\beta} {\tanh}^{-1} (\tanh (\beta ) \tanh (\beta h))\right]
\end{eqnarray}
where $P(h)$ is the average probability distribution of effective fields acting on the sites and
$Q(u)$ is that of cavity fields.  We would like to stress the importance of the fact that the strong
inhomogeneities present in the graph are correctly taken into account and handled via the computation
of the whole probability distributions.  In the Ising case we can easily work only with the cavity
fields, whose self consistent equation for the $Q(u)$ reads:
\begin{equation}
Q(u) = \frac{1}{<k>}  \sum_k k p_k  \int \prod_{t=1}^{k-1} d u_t
Q(u_t) \delta \left[u - \frac{1}{\beta} {\tanh}^{-1} \left[\tanh (\beta ) \tanh
\left(\beta \sum_t^{k-1} u_t + \beta H_0\right)\right]\right]
\label{Qdiu}  
\end{equation}
This is an integral equation that can be solved at every value of $\beta$ using a population dynamics
algorithm such as the RS simple version of that proposed in \cite{MEPA}.  The equation for the physical
magnetization probability distribution will be:
\begin{equation}
\Pi (s) = \sum_k p_k \int \prod_{t=1}^{k} d u_t Q(u_t) \delta \left[s -
\tanh \left(\beta \sum_t u_t + \beta H_0\right)\right]
\label{Pdim}
\end{equation}
The equations for $<u>$ and $<s>$ follow:
\begin{eqnarray}
<u> &=& \int u Q(u) d u = \frac{1}{<k>}  \sum_k k p_k  \int
\prod_{t=1}^{k-1} d u_t Q(u_t) \frac{1}{\beta} {\tanh}^{-1} \left[\tanh
(\beta ) \tanh \left(\beta\sum_t^{k-1} u_t +\beta H_0\right)\right] \label{umedia} \\ <s> &=&
\int s \Pi (s) d s = \sum_k p_k  \int \prod_{t=1}^{k} d u_t Q(u_t)
\tanh \left(\beta\sum_t^{k} u_t + \beta H_0\right)
\label{mmedia}
\end{eqnarray}
Plugging equations (\ref{eff1}) and (\ref{eff2}) into (\ref{eq:potenziale}) one
ends up with the following expression for the free energy:
\begin{eqnarray}
\beta F &=&  <k> \int \int dhduP(h)Q(u) \log (1 + \tanh (\beta
h) \tanh (\beta u)) - \frac{<k>}{2} \int
\prod_{t=1}^{2} dh_t P(h_t) \log (1 + \tanh (\beta)
\prod_{t=1}^{2} \tanh (\beta h_t)) \nonumber \\ & & - \sum_k p_k
\int \prod_{t=1}^{k} du_t Q(u_t) \log \left( 
\frac{2 \cosh (\beta \sum_{t=1}^k u_t + \beta H_0)}{\prod_{t=1}^{k}2 \cosh (\beta u_t)}  \right)
-<k> \log (2) + \frac{<k>}{2}(\beta - \log (\cosh (\beta )))
\label{eq:libera}
\end{eqnarray}
At the saddle point the expression further simplifies.
The internal energy and the specific heat at the saddle point can be calculated from Eq.(\ref{eq:libera})
through relations $<E> = \beta \frac{\partial \beta F}{\partial \beta}$ and 
$C = d<E>/dT$ and further exploiting (\ref{eq:saddle1a}) and (\ref{eq:saddle1b}).
The constant $<k>/2$ gauges the value of the energy to zero at $T=0$ and no external field.

\subsection{Ferromagnetic phase transition}

At $T=0$ and in the limit of non vanishing fields ($u$ and $h$ $\sim O(1)$)
it is straightforward to see that the cavity fields can take only 0 or 1
values.  The equation (\ref{Qdiu}) can be solved exactly with the ansatz $Q(u)
= q_0 \delta(u) + (1 - q_0)\delta(u-1)$. Plugging this ansatz into Eqs.
(\ref{Qdiu}), (\ref{umedia}), and (\ref{mmedia}) one obtains:
\begin{equation}
\left<u\right> = 1 - q_0, 
\label{spT0__}
\end{equation}
\begin{equation}
\left<s\right> = 1 - G_0(q_0), 
\label{spT0_}
\end{equation}
\begin{equation}
q_0 = G_1\left(q_0\right),
\label{spT0}
\end{equation}
where
\begin{equation}
G_0(x) = \sum_k k \; p_k \; x^k,\ \ \ \
G_1(x) = \frac{1}{<k>}  \sum_k k \; p_k \; x^{k-1},
\label{gfunctions}
\end{equation}
are the generating functions of the degree distributions of a vertex chosen at random and a vertex
arrived following an edge chosen at random \cite{Newman}, respectively.  We point out that these
equations correctly coincides with that obtained in the problem of percolation in a random graph with
an arbitrary degree distribution \cite{MO-RE,Newman}, where the average magnetization $\left<s\right>$
is just the size of the giant component. Moreover, these expressions can be easily generalized to
higher order hypergraphs as it has been done in \cite{RIWEZE,2p,lungo}. From Eq. (\ref{spT0_}) it
follows that there is a finite magnetization whenever the solution $q_0$ of Eq. (\ref{spT0}) is less
than 1. This happens whenever
\begin{equation}
\frac{\left<k^2\right>}{\left<k\right>}\geq2,
\label{conditionT0} 
\end{equation}
that is just the condition for percolation in a random graph \cite{MO-RE,Newman}. On the contrary, for
$\left<k^2\right>/\left<k\right><2$ the magnetization (the size of giant component) is 0,
{\em i.e.} the system is in a paramagnetic state.

For random graphs satisfying the percolation condition in Eq. (\ref{conditionT0}) we are now
interested in finding the value of $\beta_c$ for the ferromagnetic transition. There are few
equivalent ways to do so. In the general case we can derive both sides of Eq.(\ref{umedia}) in $u =
0$ self consistently, obtaining
\begin{equation} 
\frac{1}{T_c}=\beta_c =
-\frac{1}{2}\log\left(1-2\frac{\left<k\right>}{\left<k^2\right>}\right).
\label{betac2} 
\end{equation} 
In the limit $\left<k^2\right>\gg2\left<k\right>$ we can expand the logarithm getting the first order
condition $T_c=\left<k^2\right>/\left<k\right>$ which is the value found in the naive mean field
approximation (\ref{eq:5}). Hence, the MF approach in developed in the previous section is valid for
$\left<k^2\right>\gg2\left<k\right>$ and, in this case, it gives the same results as those obtained
using the replica approach.

\subsection{Critical behavior around $\beta_c$}
\label{sec:around}

The critical behavior of the thermodynamical quantities $<s>$, $\chi$, $\delta C$, and $<s>_{H_0}
\sim H_0^{1/\delta}$ close to $\beta_c$ can be calculated without having to explicitly solve the self
consistent equations for the whole probability distributions $Q(u)$ and $\Pi (s)$. Sufficiently close
to the critical point we can assume $Q(u) \sim \delta (u - <u>)$ being $<u>$ infinitesimal. In fact
this ansatz is incorrect if $\beta > \beta_c$, because it correctly takes into account the connectivity
distribution but disregards the non trivial structure of the $Q(u)$ , which does not merely
translate from the critical form $\delta (u)$ at $\beta_c$, but immediately develops a continuum
structure. In the zero temperature limit the continuum shape will again collapse in a
distribution of delta peaks discussed above.  Nevertheless, sufficiently close to the transition we
can expect only the first momenta of the $Q(u)$ to be relevant. For distributions with $<k^4>$ finite
one is left with a closed system of equations for the first three momenta all contributing to the same
leading order. Defining $\mu_n = <k(k-1)...(k-n)>$ and $A = ((\tanh (\beta))^2 \mu_2)/(\beta^2 <k> 
- (\tanh (\beta))^2 \mu_1)$
\begin{eqnarray}
<u> &=& \frac{\tanh (\beta )}{\tanh (\beta_c)}<u> -\frac{\beta^2 \tanh (\beta ) 
[1 - (\tanh (\beta )]^2}{3<k>} \left[ \mu_1<u^3> + 3\mu_2<u><u^2> + \mu_3<u>^3  \right] \nonumber \\
<u^2> &=& A <u>^2 \nonumber \\
<u^3> &=& \left( \frac{(\tanh (\beta ))^3 A \mu_2 + \mu_3}{\beta^3 <k> - 
(\tanh (\beta ))^3 <\mu_1}\right) <u>^3
\label{momenta}
\end{eqnarray}
Similar calculations can be done for the the free energy, the energy and the specific heat.
Proportionality is found also for $<k^4>=\infty$, where the calculation is a bit more involved
because the leading momenta are to be found via an analytic continuation in the values of their
order.  Correctly taking into consideration the values of the leading momenta is important in case
one is interested not only on calculating critical exponents, but also the amplitudes, because in
general more terms at the same leading order are present, as we see in Eq.(\ref{momenta}). However, the
exponents are determined by the lowest non trivial last analytic value of the momenta of the distribution
$p_k$, and do not change in the general case because all relevant momenta of the $Q(u)$ give the same
divergence in the momenta of the $p_k$. One example again is given in Eq.(\ref{momenta}). Since we
are not interested in the calculation of amplitudes we can therefore resort to the variational ansatz
$Q(u) \sim \delta (u - <u>)$ in the proximity of the transition. However we would like to stress
that calculations can be done also in the general case. Eqs.(\ref{umedia}), (\ref{mmedia}) then become
\begin{eqnarray}
<u> &\sim& \frac{1}{<k>}  \sum_k k p_k \frac{1}{\beta} {\tanh}^{-1}
(\tanh (\beta )  \tanh (\beta (k-1)<u> + \beta H_0))
\label{umediacritica} \\
<s> &\sim& \sum_k p_k \tanh (\beta k <u> + \beta H_0)
\label{umediacritica1} 
\end{eqnarray}
The corresponding expressions for the free energy, the energy and the specific heat
can be retrieved in the same way and will not be written here for the sake of space.
If $<k^4>$ is finite the first non trivial term of the power series expansion
of Eq.(\ref{umediacritica}) that still gives an analytic contribution
is simply $<u>^3$. One finds
\begin{eqnarray}
<u> &\sim& \left( \frac{3 <k>}{\beta_c^2 (\tanh \beta_c) <k (k-1)^3>
} \right)^{\frac{1}{2}} \tau^{\frac{1}{2}} \label{final} \\
<s> &\sim& <u> , \ \ \  \chi \sim \tau^{-1}, \ \ \  <s> \sim H_0^{1/3}
\label{eq:criticalmean}
\end{eqnarray}
where $\tau=1-T/T_c$ as usually defined. All exponents are the usual mean field ones. However, 
one finds a finite jump in the specific heat. The transition is therefore
first order in the traditional sense.
If we keep all the relevant momenta in our calculation, we find the expected correction to the
amplitudes. For example we find
$<u> \sim \sqrt{3}((\beta_c \tanh (\beta_c ) <k>)((\mu_1 + 3\mu_2)A + \mu_3>))^{-\frac{1}{2}}
\tau^{\frac{1}{2}}$. 
This equation reduces to (\ref{final}) if we disregard higher momenta.

\section{Power law distributed graphs}

\begin{center}
\begin{figure}
\epsfxsize=0.3\textwidth  \epsffile{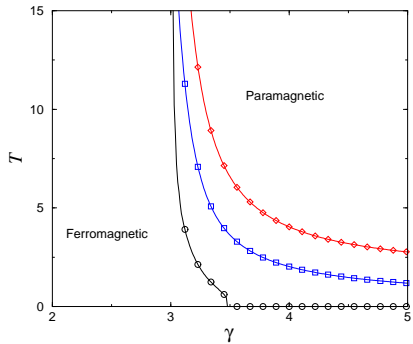}

\caption{The phase diagram of the Ising model on scale-free graphs with a power law degree
distribution $p_k=ck^{-\gamma}$, $m\leq k<\infty$. The ferromagnetic transition lines depends on the
value of $m$, with $m=1$ circles, 2 squares, and 3 diamonds. }

\label{mT0}
\end{figure}
\end{center}

In the following we are mostly interested in the case of a power law distribution of the type
\begin{equation}
p_k = c \; k^{-\gamma},\ \ \ \ m\leq k <\infty,
\label{Power}
\end{equation}
where $c$ is a normalization constant and $m$ is the lowest degree. Note that in the case of a power
law distribution
\begin{equation}
<k^2> = c \sum_{k=m}^{k_{max}} k^{2-\gamma} > c \; m \sum_{k=m}^{k_{max}}
k^{1-\gamma} = m <k>.
\label{Disug}
\end{equation}
Hence, we have that for $m\geq2$ the graph is always percolating for all $\gamma$ independently on
the cutoff $k_{max}$. Then for $m\geq2$ the critical temperature is always given by Eq.
(\ref{betac2}). In particular for $m>2$ and $\gamma\gg1$ the critical temperature approaches the
limit $T_c^{lim}=-2/\log(1-2/m)$ while for $m=2$ the critical temperature tends to zero in the large
$\gamma$ limit. However, for $m = 1$ there is a critical value $\gamma^{\star}$ beyond which the
graph is no longer percolating \cite{Chung}. $\gamma^{\star}$ is the value of $\gamma$ at which
$\left<k^2\right>=2\left<k\right>$, resulting $\frac{\zeta(\gamma^{\star} - 2)}{\zeta(\gamma^{\star}
- 1)} = 2$ that have the solution $\gamma^{\star} = 3.47875...$. If $\gamma \ge \gamma^{\star}$ the
system is always paramagnetic while for $\gamma<\gamma^{\star}$ there is a transition to a
ferromagnetic state at a temperature given by Eq. (\ref{betac2}). In Fig.~\ref{mT0} we show the phase
diagram together with the critical lines for $m=1$, 2 and 3.

\subsection{$2<\gamma\leq3$}

For $2<\gamma\leq3$ the second moment of the degree distribution diverges and, therefore, as
discussed in previous sections, the system is always in a ferromagnetic state. In this case it is
important to investigate the behavior of $\left<u\right>$ and $\left<s\right>$ when
$\beta\rightarrow0$. This computation can be done using either the mean-field or the replica approach
obtaining the same results. In fact, in this case we have $\lim_{\beta \to 0} Q(u) = \delta
(u)$ and putting this limit distribution into the self consistent equation for $<u>$ and $<s>$ we
recover the mean field asymptotic behavior. For $2<\gamma\leq3$ the sums in Eq.(\ref{eq:4}) are
dominated by the large $k$ region. In this case these sums can be approximated by
integrals resulting
\begin{equation}
\left<u\right>\approx(\gamma-2)(m\beta \left<u\right>)^{\gamma-2}
\int_{m\beta \left<u\right>}^\infty dxx^{1-\gamma}\tanh x,
\label{eq:7}
\end{equation}
while the magnetization, $\left<s\right>=\sum_kp_k\tanh\left(\beta
k\left<u\right>\right)$, is simply given by
\begin{equation}
\left<s\right>\approx\frac{\gamma-1}{\gamma-2}m\beta 
\left<u\right>.
\label{eq:8}
\end{equation}
The above integral cannot be analytically calculated but its asymptotic behaviors for
$\beta\rightarrow0$ can be obtained. For $\gamma=3$ the integral in the rhs of Eq. (\ref{eq:7})  is
dominated by the small $x$ behavior. Thus, approximating the $\tanh x$ by $x$ and computing
$\left<u\right>$ we obtain
\begin{equation}
\left<u\right>\approx\frac{\exp(-1/m\beta)}{m\beta},\ \ \ \ \gamma=3.
\label{eq:9}
\end{equation}
On the other hand, for $\gamma<3$ the integral in the rhs of Eq. (\ref{eq:7}) is finite for any value
of $m\beta \left<u\right>$ and, therefore, for $m\beta \left<u\right>\ll1$ it follows that
\begin{equation}
\left<u\right>\approx[(\gamma-2)I]^{\frac{1}{3-\gamma}}
(m\beta)^{\frac{\gamma-2}{3-\gamma}},
\ \ \ \ \gamma<3,
\label{eq:10}
\end{equation}
where $I=\int_0^\infty dxx^{1-\gamma}\tanh x$.  Finally, substituting eqs. (\ref{eq:9}) and
(\ref{eq:10}) on Eq. (\ref{eq:8}) we get
\begin{equation}
\left<s\right>\sim\exp(-1/m\beta),\ \ \ \ \gamma=3,
\label{eq:11}
\end{equation}
\begin{equation}
\left<s\right>\sim(m\beta)^{\frac{1}{3-\gamma}},\ \ \ \ 2<\gamma<3.
\label{eq:12}
\end{equation}
With the same technique one can study the behavior of the other physically relevant
quantities. Extracting the leading asymptotic terms from the expressions for
the energy and the specific heat $T = \infty$, 
we find an infinite order phase transition with
\begin{eqnarray}
\delta C &\sim& \frac{e^{-2/m\beta}}{\beta^2}, \ \ \ \ \gamma=3, \nonumber \\
\delta C &\sim& \beta^{(\gamma-1)(3-\gamma)}, \ \ \ \ 2<\gamma<3.
\label{csp}
\end{eqnarray}
Extracting from Eq.(\ref{umedia}) the leading behavior 
of $\chi_u \equiv \partial <u>/\partial H_0$ at $H_0 = 0$ and plugging the result
together with eqs.(\ref{eq:9}), (\ref{eq:10}), (\ref{eq:11}) and (\ref{eq:12}) into 
$\chi \equiv \partial <m>/\partial H_0$
one obtains:
\begin{equation}
\chi \sim \frac{1}{m^2 \beta}
\label{chi}
\end{equation}

The limiting case $\gamma=3$ corresponds with the Barabasi-Albert model studied in
\cite{aleksiejuk01} by means of numerical simulations.  The magnetization exhibits an exponential
decay in agreement with our calculation in Eq. (\ref{eq:11}). Moreover, the critical temperature was
observed to increase logarithmically with the network size $N$. Computing $T_c$ in Eq. (\ref{eq:5})
for $\gamma=3$ we obtain $T_c\approx (m/2)\ln N$, which is in very good agreement with their
numerical results. It is worth remarking that similar exponential and logarithmic dependencies have
been observed for the order and control parameter in some non-equilibrium transitions
\cite{vespignani01,moreno01}.

\subsection{$3<\gamma\leq5$}

In this case $\left<k^2\right>$ is finite and, therefore, there is a ferromagnetic transition
temperature given by Eq. (\ref{betac2}). However, $\left<k^4\right>$ is not finite and the
derivation of the MF critical exponents performed in Sec. \ref{sec:around} is not valid.  In order to
find the critical exponents we can write the functions inside the connectivities sums as power series
in $<u>$. The coefficients of the two series will however depend on the higher momenta of the
connectivities distribution and will be infinite beyond a certain power of $<u>$. This is direct
consequence of the fact that the power expansion of the $\tanh (y)$ around 0 is convergent as long as
the $y < \pi/2$, while for any $<u>$ in our cases there will exist an $k^{\star}$ such that $<u>$
lays outside the convergence radius. Nevertheless, the function is well approximated by the expansion
when one truncates it up to the maximum analytical value of the exponent such that all momenta of the
power law distribution taken into consideration are finite.

For $3<\gamma<5$ the highest analytical exponent of the expansion of Eq.(\ref{umediacritica}) in
powers of $<u>$ is $n_{max} = \gamma - 2$, where the integer value has been analytically continued
and so should be done with the corresponding series coefficient. In this range of values of s
$n_{max}$ is lower than 3 so is to be taken as the correct value instead of $n = 3$ that leads to non
analicities. With analogous calculations we are able to find all other critical exponents:
\begin{eqnarray}
<u> &\sim& <s> \sim \tau^{\frac{1}{\gamma - 3}} \nonumber \\
\delta C &\sim& \tau^{(5 - \gamma)/(\gamma - 3)} \nonumber \\
\chi &\sim& \tau^{-1}, \ \ \  <s> \sim H_0^{\beta/(1 + \beta)} \sim H_0^{1/(\gamma -2)}
\label{eq:critical35}
\end{eqnarray}
On the other hand, for $\gamma=5$ one can find a logarithmic correction to the previous values
expanding the inverse hyperbolic tangent in Eq.(\ref{umediacritica}) to the third order in the tails
of the connectivities distribution. The results are:
\begin{eqnarray}
<u> &\sim& <s> \sim \tau^{1/2}/(-\log(\tau ))^{1/2} \nonumber \\
\delta C &\sim& 1/(-\log(\tau )) \nonumber \\
\chi &\sim& \tau^{-1}, \ \ \  <s> \sim H_0^{1/3}/(-\log(H_0))^{1/3}
\label{eq:critical5}
\end{eqnarray}
The specific heat is continuous at $\beta_c$ for $\gamma \in (3,5]$, indicating a 
phase transition of order $>$ 1. A part from the logarithmic corrections in the $\gamma = 5$
case, the universality relations between the exponents are satisfied.

This treatment parallels the $T=0$ calculations done in \cite{CAH} for the case of percolation
critical exponents in a power law graph in presence of further dilution. If we introduce a cutoff
into the connectivities distribution the critical exponent very close to the transition point is
always the mean field one, due to the fact that the sum over the connectivities is always finite and
there is no non analyticity in $<u> = 0$ for any $\gamma$.  However the influence of non trivial
terms is very strong (decreasing if we increase $\gamma$). Eq.(\ref{final}) is always valid but only
in a very narrow region around $\beta_c$. The numerical values of $T_c$ and of the amplitudes in the
critical behavior of the magnetization are also strongly affected being a function of the moments of
the connectivities distribution. In the infinite cutoff limit the mean field window shrinks to zero
and one recovers the non trivial behavior. Indeed, if we work with large enough a cutoff at $\gamma
\in (3,5)$ and calculate the average magnetization in regions where $\beta (k-1) <u>(\beta) \sim
\pi/2$, limit of the radius of convergence of the series expansions of ${\tanh}^{-1} (\tanh (\beta )  
\tanh (\beta (k-1)<u>))$, we see a contribution in the magnetization curves that goes as $(\beta -
\beta_c)^{1/(\gamma-3)}$. This region becomes dominant for large values of the cutoff.

\section{Summary and conclusions}

In summary, we have obtained the phase diagram of the Ising model on a random graph with an arbitrary
degree distribution. Three different regimes are observed depending on the moments $\left<k^2\right>$
and $\left<k^4\right>$ of the distribution. For $\left<k^4\right>$ finite the critical exponents of
the ferromagnetic phase transition coincides with those obtained from the simple MF theory. On the
contrary, for $\left<k^4\right>$ not finite but $\left<k^2\right>$ finite we found non-trivial
exponents that depend on the power law exponent of the degree distribution $\gamma$. On the other
hand, for $\left<k^2\right>$ not finite the system is always in a ferromagnetic state. Moreover, at
$T=0$ we recover the results obtained by the generating function formalism for the 
percolation problem on random graphs with an arbitrary degree distribution.

\end{document}